\documentclass[%
 reprint,
superscriptaddress,
showpacs,
 amsmath,amssymb,
pra,
]{revtex4-1}

\usepackage{graphicx}
\usepackage{dcolumn}
\usepackage{bm}
\usepackage{color}
\usepackage{hyperref}
\hypersetup{backref,colorlinks=true,breaklinks,urlcolor=blue,linkcolor=blue,citecolor=blue}
\usepackage{siunitx}

\usepackage[section]{placeins}
\usepackage{longtable}
\usepackage{array}
\usepackage{float}
\usepackage{colortbl}
\definecolor{mygray}{gray}{0.96}

\begin{document}

\preprint{APS/123-QED}

\title{Broad Feshbach resonances in ultracold alkali-metal systems}
\author{Yue Cui}
\author{Min Deng}
\affiliation{State Key Laboratory of Low Dimensional Quantum Physics, Department of Physics, Tsinghua University, Beijing 100084, China}
\author{Li You}
\affiliation{State Key Laboratory of Low Dimensional Quantum Physics, Department of Physics, Tsinghua University, Beijing 100084, China}
\affiliation{Collaborative Innovation Center of Quantum Matter, Beijing, China}
\author{Bo Gao}
\email{bo.gao@utoledo.edu}
\affiliation{Department of Physics and Astronomy, University of Toledo, Mailstop 111, Toledo, Ohio 43606, USA}
\author{Meng Khoon Tey}
\email{mengkhoon\_tey@mail.tsinghua.edu.cn}
\affiliation{State Key Laboratory of Low Dimensional Quantum Physics, Department of Physics, Tsinghua University, Beijing 100084, China}
\affiliation{Collaborative Innovation Center of Quantum Matter, Beijing, China}
\pacs{34.10.+x, 34.50.Cx, 67.85.-d, 67.60.Bc}

\date{\today}

\begin{abstract}

A comprehensive search for ``broad'' Feshbach resonances (FRs) in all possible combinations of stable alkali-metal atoms is carried out, using a multi-channel quantum-defect theory assisted by the analytic wave functions for a long-range van-der-Waals potential. A number of new ``broad" $s$-, $p$- and $d$-wave FRs in the lowest-energy scattering channels, which are stable against two-body dipolar spin-flip loss, are predicted and characterized. Our results also show that ``broad" FRs of $p$- or $d$-wave type that are free of two-body loss do not exist between fermionic alkali-metal atoms for magnetic field up to 1000\,G. These findings constitute helpful guidance on efforts towards experimental study of high-partial-wave coupling induced many-body physics.

\end{abstract}

\maketitle

\section{Introduction}
Feshbach resonance (FR) enables versatile tuning of effective interactions between ultracold atoms~\cite{Chin2010FRreview}. It has facilitated a wealth of interesting studies on few- and many-body physics using ultracold quantum gases. Of particular interests to experiments are ``broad" FRs with open-channel dominated characteristics~\cite{Chin2010FRreview}, where the effective interaction between atoms can be modeled as a single-channel scattering problem and the long-range van-der-Waals universality applies~\cite{Gao2011MQDTFR}.

Capitalizing on the ``broad''~\cite{Chin2010FRreview,Gao2011MQDTFR} $s$-wave FRs in fermionic $^6$Li and $^{40}$K, which feature small two-body and three-body collision losses, tremendous advances have been achieved in the study of BEC-BCS crossover and the unitary Fermi gas ~\cite{Jin2004BECBCS,Grimm2004crossover,Ketterle2004FRcondensation,Kinast2005unitaryFG,Horikoshi2010unitaryFG,Nascimbene2010unitaryFG,Ku2012EOS,Sidorenkov2013Secondsound,Tey2013CollectiveModes}. The study of atomic gases near ``broad'' FRs of nonzero partial waves, on the other hand, has only witnessed limited activities, although they are predicted to exhibit richer physics accompanied by quantum phases nonexistent with s-wave interactions ~\cite{Botelho2005,Andreev2005pwavetransition,Yip2005FermiSuperfluid,Gurarie2007pwavesuperfluid,Radzihovsky2009pwave}. This dichotomy is mainly due to the fact that ``broad'' high-partial-wave FRs which are stable against collision losses have yet to be found. In fact, ``broad" nonzero-partial-wave FRs have so far only been reported in $^{85}$Rb-$^{87}$Rb mixtures~\cite{pappthessis,You2016Rb85Rb87,You2017dwave} and $^{41}$K atoms~\cite{Yao2017dwave}. However, FRs studied in these systems exhibit substantial collision losses due to the bosonic nature of the atoms.


In this paper, we summarize the main results from an extensive search of ``broad" $s$-, $p$- and $d$-wave FRs in all possible combinations of stable alkali-metal atoms. In particular, we focus on FRs in the lowest-energy scattering channels where the exothermic two-body dipolar spin-flip collision cannot occur. The search is carried out using an analytic multi-channel quantum-defect theory (MQDT)~\cite{Gao1998r6solutions,Gao2005MQDT,Gao2009singlechannel,Gao2011MQDTFR}. This theory provides the simplest description of magnetic FRs in alkali-metal interactions, using only three parameters for all partial waves. This simplicity and efficiency make it ideal for exploring trends and qualitative features in a large number of systems.


The main results of the current work can be summarized as follows: First, new ``broad" $s$-, $p$- and $d$-wave resonances are predicted in many alkali-metal systems. The systems that exhibit rich spectra of ``broad" nonzero partial-wave FRs include pure $^{41}$K gas, and $^{41}$K-$^{87}$Rb, $^{39}$K-$^{133}$Cs, $^{41}$K-$^{133}$Cs, $^{85}$Rb-$^{87}$Rb, $^{85}$Rb-$^{133}$Cs mixtures. Second, ``broad" $p$- and $d$-wave FRs free of two-body loss are not found in all fermionic alkali-metal gases (including their mixtures) for magnetic field up to 1000\,G. Third, the $^{41}$K gas contains a unique and extraordinarily ``broad" $d$-wave shape resonance~\cite{Yao2017dwave}. Due to its real single-channel feature, this resonance could be of particular interest for studying $d$-wave interaction induced many-body physics.

The paper is organized as follows. Section~\ref{section2} presents the main results of our search, with Sec.~\ref{subsection1} reporting the FRs for the fermionic alkali-metal gases, and Sec.~\ref{subsection2} for the ``broad" resonances of all other alkali-metal systems, including those between bosonic atomic species, as well as between fermionic and bosonic species. The ``broad'' $d$-wave shape resonance of $^{41}$K, which was first studied in Ref.~\cite{Yao2017dwave}, is highlighted and discussed in detail in Sec.~\ref{subsection3}. Section~\ref{section3} briefly describes the theoretical model adopted for our study. It includes the essential computation details needed for obtaining the results presented in this work. Sec. \ref{section4} concludes this work. Appendixes~\ref{appendixA} and \ref{appendixB} contain computation details and tabulate the optimized parameters we used for all alkali-metal systems.

\section{Main Results}\label{section2}

This section reports the FRs predicted using our model for all combinations of stable alkali-metal atoms. As our main motivation is to find low-loss FRs that are potentially useful for the study of many-body physics, we focus only on the lowest-energy scattering channels (open channels) where exothermic two-body dipolar spin-flip collisions are forbidden. Exceptions are made for pure homonuclear fermionic gases, where we include both the two lowest-energy scattering channels since two-body relaxations cannot occur in the excited one due to the Pauli exclusion principle.

Furthermore, since we aim mainly at ``broad'' FRs which arise from the strong electronic (Coulomb) coupling between the open channel and the closed channel, our model ignores the weak anisotropic magnetic dipole-dipole~\cite{Verhaar1988dipole,Axelsson1995dipole} and second order spin-orbit~\cite{Krauss1996spinorbit,Julienne2000spinorbit} interactions between the atoms. Consequently, all supposedly ``narrow'' FRs arising from the coupling between open and closed channels with different partial waves or with different $m_{f_1}+m_{f_2}$ would be missed out from our searches [$m_{f_i}$ is the azimuthal spin of the magnetic-field-dressed hyperfine state $|f_{i}m_{f_i}\rangle$ $(i=1,2)$ (Eq.\,(\ref{dressedhyperfinestate})) of the two colliding atoms].

We characterize every predicted resonance by its position $B_{0l}$, resonance width $\Delta_{Bl}$, normalized background scattering length $\tilde{a}_{\textmd{bgl}}$, differential magnetic moment between the molecular and atomic state $\delta\mu_{l}$, experimentally measured position $B_{0l}^{\textmd{expt}}$ (if observed), and the resonance strength parameter $\zeta_{\textmd{res}}$. The last parameter is of most interest since it indicates whether a resonance is ``broad" ($|\zeta_{\textmd{res}}|\gg1$) or ``narrow" ($|\zeta_{\textmd{res}}|\ll1$)~\cite{Gao2011MQDTFR}. The detailed definitions of the aforementioned parameters follow from the notations of Gao~\cite{Gao2011MQDTFR}, and can be found in Sec.~\ref{section3}.

The accuracies of our predictions for the resonance positions are typically within a few percent, which are not as high as those obtained using coupled-channel (CC) calculations based on full molecular potentials. Nevertheless, it should be largely sufficient for identifying systems and resonances of interest. For any particular system and/or resonance, the description can be further refined when necessary.

\subsection{FRs in fermionic alkali-metal systems}\label{subsection1}

FRs between fermionic atoms play a special role in the experimental study of ultracold gases, not only because Fermi statistics of the atoms are essential for quantum simulation studies of condensed matter models, but also because Fermi statistics help to suppress three-body and many-body recombination~\cite{Petrov2004fermiloss}, and to form long-lived strongly interacting systems. The stable strongly interacting Fermi gas near the 832-G ``broad'' $s$-wave resonance of $^6$Li is perhaps the most celebrated example. By the same token, it is highly desirable to find ``broad'' high-partial-wave FRs in fermionic systems.

The only stable fermionic alkali-metal atoms are $^{6}$Li and $^{40}$K. Table~\ref{table0} presents all $s$-, $p$-, and $d$-wave FRs free from two-body losses, predicted using our model for $^{6}$Li-$^{6}$Li, $^{40}$K-$^{40}$K and $^{6}$Li-$^{40}$K within the magnetic-field range from 0 to 1000\,G. Except for the well-known ``broad'' $s$-wave FRs at 832\,G in $^{6}$Li$|\frac{1}{2}, \frac{1}{2}\rangle$ + $^{6}$Li$|\frac{1}{2}, -\frac{1}{2}\rangle$ ($\zeta_{\textmd{res}}=142$) and that at  200\,G in $^{40}$K$|\frac{9}{2}, -\frac{9}{2}\rangle$ + $^{40}$K$|\frac{9}{2}, -\frac{7}{2}\rangle$ ($\zeta_{\textmd{res}}=3.1$), all other resonances are found to be very ``narrow" ($|\zeta_{\textmd{res}}|\ll1$).

\begin{table*}[t]
\renewcommand\arraystretch{1.5}
\caption{$s$-, $p$-, and $d$-wave FRs free from two-body loss in the fermionic $^{6}$Li-$^{6}$Li, $^{40}$K-$^{40}$K, and $^{6}$Li-$^{40}$K systems, for magnetic field in the range of 0-1000\,G.  The scattering channels of the systems are labeled in magnetic-field-dressed hyperfine basis $|f,m_{f}\rangle$ (Eq.\,(\ref{dressedhyperfinestate})).}

\begin{tabular}{|p{3.8cm}|p{0.7cm}<{\centering}|p{1.4cm}<{\centering}|p{1.4cm}<{\centering}|p{1.4cm}<{\centering}|p{1.4cm}<{\centering}|p{1.4cm}<{\centering}|p{1.6cm}<{\centering}|p{2.6cm}<{\centering}|}

\hline
Scattering channel     &   $l$   &   $B_{0l}$(G) &    $\zeta_{\textmd{res}}$ &   $\Delta_{Bl}$ (G)   &  $\tilde{a}_{\textmd{bgl}}/\bar{a}_{l} $   &   $\delta\mu_{l}/\mu_{B}$ &   $B_{0l}^{\textmd{expt}}$(G) & Remark/Reference \\
\hline
$^{6}$Li$|\frac{1}{2}, \frac{1}{2}\rangle$ + $^{6}$Li$|\frac{1}{2}, \frac{1}{2}\rangle$  &     $p$     &   159.2  &   -0.22   &  -44.4  &   -1.2  &   2.1  &   159.14   &  ~\cite{Salomon2004pwaveLi,Ketterle2005Li6}   \\
  $^{6}$Li$|\frac{1}{2}, \frac{1}{2}\rangle$ + $^{6}$Li$|\frac{1}{2}, -\frac{1}{2}\rangle$  &      $s$    &   554.4  &      0.0012 &   0.098   &  2.0  &  2.0 &   543.25  &   ~\cite{Hulet2003Li6}   \\
  &     $s$    &   832.5  &    142.0  &  -293.6   &  -40.6  &   3.6  &   832.18  &   ~\cite{Jochim2013Li6}  \\
  &    $p$     &   185.2   &   -0.14 & -26.3  &  -1.5   &   2.0  &  185.09  &   ~\cite{Salomon2004pwaveLi,Ketterle2005Li6}  \\
  &     $d$   &   -   &   -   &    - &    -&    -&   & not found \\
\rowcolor{mygray}[0.6\tabcolsep]
$^{40}$K$|\frac{9}{2}, -\frac{9}{2}\rangle$ + $^{40}$K$|\frac{9}{2}, -\frac{9}{2}\rangle$ &     $p$   &   -   &   -   &    - &    -&    -&    - & not found \\ \rowcolor{mygray}[0.6\tabcolsep]
$^{40}$K$|\frac{9}{2}, -\frac{9}{2}\rangle$ + $^{40}$K$|\frac{9}{2}, -\frac{7}{2}\rangle$ &    $s$     &  200.3   &      3.1  & 8.0 &  2.7  &  1.7  &   202.1  &   ~\cite{Jin2004BECBCS} \\ \rowcolor{mygray}[0.6\tabcolsep]
  &     $p$   &   -   &   -   &    -  &    - &    - &    - & not found \\ \rowcolor{mygray}[0.6\tabcolsep]
  &    $d$    &  60.4   &       -0.25  & 30.1 &  0.42   &  1.6  &  -  &  -  \\
$^{6}$Li$|\frac{1}{2}, \frac{1}{2}\rangle$ + $^{40}$K$|\frac{9}{2}, -\frac{9}{2}\rangle$  &   $s$     &   157.6  &   0.0035 & 0.14  &  1.6  &  1.7  &  157.6  &   ~\cite{Grimm2008LiK,Gao2014LiKFR}  \\
  &     $s$     &   167.6  &  0.0028  &  0.11  &  1.6  &  1.8  & 168.2  &   ~\cite{Grimm2008LiK,Gao2014LiKFR} \\
  &     $p$   &   247.1   & -0.0013   & 0.45  &3.4  &  0.15  &  249  &   ~\cite{Grimm2008LiK,Gao2014LiKFR}  \\
   &     $d$   &   -   &   -   &    -   &    - &    - &    - &  not found \\
\hline
\end{tabular}
\\

\label{table0}
\end{table*}

\subsection{FRs in all other alkali-metal systems}\label{subsection2}

In this subsection, all ``broad" $s$-, $p$-, and $d$-wave FRs in the lowest-energy scattering channels for all other alkali-metal systems are presented. Taking the possible uncertainty of $\zeta_{\textmd{res}}$ (due to the approximations adopted in our model) into consideration, we extend the range of listed ``broad" FRs to $|\zeta_{\textmd{res}}|\geq0.5$. Table~\ref{table1} shows the results for homonuclear alkali-metal systems. ``Broad" $d$-wave resonances are found in bosonic $^{41}$K, $^{87}$Rb and $^{133}$Cs atoms.

Table~\ref{table2},~\ref{table3},~\ref{table4} and~\ref{table5} show the results for alkali-metal mixtures Li-X (X being isotopes of Li, Na, K, Rb and Cs), Na-X (X being isotopes of K, Rb and Cs), K-X (X being isotopes of K, Rb and Cs) and Rb-X (X being isotopes of Rb and Cs), respectively. The systems that exhibit ``broad" $p$-wave resonances are $^{7}$Li-$^{41}$K, $^{7}$Li-$^{87}$Rb, $^{7}$Li-$^{133}$Cs, $^{23}$Na-$^{85}$Rb, $^{23}$Na-$^{87}$Rb, $^{39}$K-$^{41}$K, $^{39}$K-$^{87}$Rb, $^{40}$K-$^{85}$Rb, $^{41}$K-$^{85}$Rb, $^{41}$K-$^{87}$Rb, $^{39}$K-$^{133}$Cs, $^{41}$K-$^{133}$Cs, $^{85}$Rb-$^{87}$Rb, $^{85}$Rb-$^{133}$Cs. In particular, a very ``broad" $p$-wave FR is predicted in the $^{41}$K-$^{87}$Rb mixture at 850.8\,G with $|\zeta_{\textmd{res}}|=244.8$. The systems that possess ``broad" $d$-wave resonances include $^{7}$Li-$^{133}$Cs, $^{23}$Na-$^{85}$Rb, $^{23}$Na-$^{87}$Rb, $^{23}$Na-$^{133}$Cs, $^{39}$K-$^{85}$Rb, $^{39}$K-$^{87}$Rb, $^{41}$K-$^{87}$Rb, $^{39}$K-$^{133}$Cs, $^{85}$Rb-$^{87}$Rb, $^{85}$Rb-$^{133}$Cs.

\begin{table*}[t]
\renewcommand\arraystretch{1.5}
\caption{``Broad" $s$-, $p$- and $d$-wave FRs in the lowest-energy scattering channels of the homonuclear alkali-metal systems. Resonances with $|\zeta_{\textmd{res}}|$ $\geq$ 0.5 for magnetic field in the range of 0-1000\,G are listed. Such ``broad" FRs are not found in systems of $^{6}$Li, $^{23}$Na and $^{40}$K atoms.}
\begin{tabular}{|p{3.4cm}|p{0.7cm}<{\centering}|p{1.3cm}<{\centering}|p{1.3cm}<{\centering}|p{1.3cm}<{\centering}|p{1.3cm}<{\centering}|p{1.3cm}<{\centering}|p{2.9cm}<{\centering}|p{3cm}<{\centering}|}

\hline
 Scattering Channel &  $l$  &        $B_{0l}$ (G) &   $\zeta_{\textmd{res}}$   &   $\Delta_{Bl}$ (G)   &  $\tilde{a}_{\textmd{bgl}}/\bar{a}_{l} $   &   $\delta\mu_{l}/\mu_{B}$     &   $B_{0l}^{\textmd{expt}}$ (G)  & Remark/Reference  \\
\hline
$^{7}$Li$|1, 1\rangle$ + $^{7}$Li$|1, 1\rangle$ &     $s$     &   738.9   &   0.88  &   -182.0  &   -0.55  &   2.1   &   -  &   -  \\
\rowcolor{mygray}[0.6\tabcolsep]
$^{39}$K$|1, 1\rangle$ + $^{39}$K$|1, 1\rangle$   &     $s$  &   402.4  &   3.7    &   -54.4 &   -0.43  &   2.0   &   403.4  &  ~\cite{Simoni2007K39}   \\
$^{41}$K$|1, 1\rangle$ + $^{41}$K$|1, 1\rangle$    &     $d$   &   17.7   &   -203.7 &   -146.5 &   10.0  &   -10.7$^{\flat}$   &   16.83/17.19/18.75$^{\sharp}$   &  shape resonance~\cite{Yao2017dwave}  \\
                                                    &     $d$  &   510.6  &  -2.6  &   5.4 &   19.0  &   2.0    &   - &   -  \\
\rowcolor{mygray}[0.6\tabcolsep]
$^{85}$Rb$|2, 2\rangle$ + $^{85}$Rb$|2, 2\rangle$    &     $s$    &   849.0  &   6.4   &   -2.1 &   -5.2  &   2.0   &   852.3  &  ~\cite{Cornish2013Rb85}  \\
$^{87}$Rb$|1, 1\rangle$ + $^{87}$Rb$|1, 1\rangle$  &     $d$   &   869.7  &   -0.9   &   -2.7 &   -2.9 &   2.8   &   930.02${^\natural}$  &  ~\cite{Verhaar2002Rb87}    \\
\rowcolor{mygray}[0.6\tabcolsep]
$^{133}$Cs$|3, 3\rangle$ + $^{133}$Cs$|3, 3\rangle$   &     $s$     &   572.4 &   456.9   &   14.6 &   24.2  &  1.9   &  549  &  ~\cite{Grimm2013Cs}           \\\rowcolor{mygray}[0.6\tabcolsep]
                                                      &     $s$     &   770.4 &  1890.8  &   113.9 &  15.5  &   1.6   &   787  &  ~\cite{Grimm2013Cs}            \\\rowcolor{mygray}[0.6\tabcolsep]
                                                      &     $d$     &   820.8 &  -1.4     &   8.9 &   0.94  &   1.7   &   820.37${^\natural}$  &  ~\cite{Grimm2013Cs}   \\

\hline
\end{tabular}
\\
\begin{flushleft}
$\flat$ The parameter $\delta\mu_{l}$ is the differential magnetic moment between the closed and open channels in a typical magnetic Feshbach resonance~\cite{Gao2011MQDTFR}. For a shape resonance where the molecular state is also supported by the open channel, $\delta\mu_{l}$ should be read as an effective parameter.

$\sharp$ The observed triplet structure of a $d$-wave FR.

$\natural$ When the triplet structure is not observed, the open channel of the FR cannot be unambiguously confirmed to be $d$ wave. The observed feature can also come from the coupling between an $s$-wave open channel to a $d$-wave closed channel~\cite{You2017dwave}.
\end{flushleft}
\label{table1}
\end{table*}

\begin{table*}[h]
\renewcommand\arraystretch{1.5}
\caption{Same as Table~\ref{table1} but for the $^{6}$Li-X and $^{7}$Li-X mixtures, X being isotopes of Li, Na, K, Rb, or Cs atom. ``Broad" $s$-, $p$-, and $d$-wave FRs in the lowest-energy scattering channels are not found in the mixtures of $^{6}$Li-$^{7}$Li, $^{6}$Li-$^{23}$Na, $^{7}$Li-$^{23}$Na, $^{6}$Li-$^{39}$K, $^{6}$Li-$^{40}$K, $^{6}$Li-$^{41}$K, $^{7}$Li-$^{40}$K, $^{6}$Li-$^{85}$Rb, or $^{6}$Li-$^{87}$Rb.}
\begin{tabular}{|p{3.8cm}|p{0.7cm}<{\centering}|p{1.3cm}<{\centering}|p{1.3cm}<{\centering}|p{1.3cm}<{\centering}|p{1.3cm}<{\centering}|p{1.3cm}<{\centering}|p{1.7cm}<{\centering}|p{2.7cm}<{\centering}|}
\hline
Scattering channel  &     $l$ &     $B_{0l}$ (G)  &   $\zeta_{\textmd{res}}$   &   $\Delta_{Bl}$ (G)   &  $\tilde{a}_{\textmd{bgl}}/\bar{a}_{l} $   &   $\delta\mu_{l}/\mu_{B}$   &  $B_{0l}^{\textmd{expt}}$ (G) &  Remark/Reference  \\
\hline
$^{6}$Li$|1/2, 1/2\rangle$ + $^{133}$Cs$|3, 3\rangle$   &    $s$      &   841.4 &   1.0   &   -57.8 &  -0.67  &   2.2    &    843.5 & ~\cite{Weidemuller2013Li6Cs133}  \\
\rowcolor{mygray}[0.6\tabcolsep]
$^{7}$Li$|1, 1\rangle$ + $^{39}$K$|1, 1\rangle$   &     $s$    &   318.8  &   1.1 &   30.0   &  2.2  &   1.5       &   -  &   - \\
$^{7}$Li$|1, 1\rangle$ + $^{41}$K$|1, 1\rangle$   &     $p$      &   765.3  &   -5.7  &   89.1 &   5.7  &   1.6    &   -  &   - \\
\rowcolor{mygray}[0.6\tabcolsep]
$^{7}$Li$|1, 1\rangle$ + $^{85}$Rb$|2, 2\rangle$  &     $s$      &   143.1  &   0.64   &   -12.6 &  -1.3  &   3.2  &   - &   - \\
$^{7}$Li$|1, 1\rangle$ + $^{87}$Rb$|1, 1\rangle$   &     $s$      &   652.9  &   5.8 &   -203.3 &  -1.4  &   2.5    &   649   &   ~\cite{Courteille2009Li7Rb87}  \\
                                                   &     $p$     &   433.5 &   -0.85  &   -32.1  &  -1.5  &   2.4   &   - &   -  \\
\rowcolor{mygray}[0.6\tabcolsep]
$^{7}$Li$|1, 1\rangle$ + $^{133}$Cs$|3, 3\rangle$  &    $p$   &   409.1  &   -2.4  &   -57.9 &  -2.1  &   2.3     &   -  &   -  \\\rowcolor{mygray}[0.6\tabcolsep]
                                                  &     $d$  &   25.4  &   -0.65 &   102.3  &  1.3  &   2.2    &   -  &   - \\

\hline
\end{tabular}
\\
\label{table2}
\end{table*}

\begin{table*}[h]
\renewcommand\arraystretch{1.5}
\caption{Same as Table~\ref{table1} but for the $^{23}$Na-X mixtures, X being isotopes of K, Rb, or Cs atom.}
\begin{tabular}{|p{4cm}|p{0.7cm}<{\centering}|p{1.3cm}<{\centering}|p{1.3cm}<{\centering}|p{1.3cm}<{\centering}|p{1.3cm}<{\centering}|p{1.3cm}<{\centering}|p{1.8cm}<{\centering}|p{2.7cm}<{\centering}|}
\hline
 Scattering channel     &  $l$ &    $B_{0l}$ (G) &   $\zeta_{\textmd{res}}$   &   $\Delta_{Bl}$ (G)   &  $\tilde{a}_{\textmd{bgl}}/\bar{a}_{l} $   &   $\delta\mu_{l}/\mu_{B}$    &         $B_{0l}^{\textmd{expt}}$ (G) &  Remark/Reference \\
\hline
$^{23}$Na$|1, 1\rangle$ + $^{39}$K$|1, 1\rangle$   &     $s$    &   441.8 &   5.6  &   -37.3 &  -1.8  &   2.0     &   -  &   -  \\
\rowcolor{mygray}[0.6\tabcolsep]
$^{23}$Na$|1, 1\rangle$ + $^{40}$K$|9/2, -9/2\rangle$  &     $s$       &   77.7 &   0.98  &   -5.6 &  -3.9  &   2.0    &   78.3   &   ~\cite{Zwierlein2012NaK}  \\\rowcolor{mygray}[0.6\tabcolsep]
                                                       &     $s$    &   88.7 &   14.0  &   -8.9 &  -20.4  &   2.3      &   88.2   &  ~\cite{Zwierlein2012NaK}  \\
$^{23}$Na$|1, 1\rangle$ + $^{41}$K$|1, 1\rangle$ &     $s$     &   73.1 &   2.3  &   4.6 &  5.1  &  2.3    &   -  &   -  \\
                                                 &     $s$    &   470.6 &   3.1 &   6.2 &  5.6  &  2.1     &   -  &   -  \\
\rowcolor{mygray}[0.6\tabcolsep]
$^{23}$Na$|1, 1\rangle$ + $^{85}$Rb$|2, 2\rangle$  &     $s$      &   314.3  &   0.83  &   5.5 &  1.5  &  1.7   &   -  &   -  \\\rowcolor{mygray}[0.6\tabcolsep]
                                                   &     $p$       &   173.8 &   -6.5   &   19.3 &  5.8  &  1.8   &   -  &   - \\\rowcolor{mygray}[0.6\tabcolsep]
                                                   &     $p$        &   219.6 &   -3.5  &   37.8 &  2.0  &  1.6   &   -&   -  \\\rowcolor{mygray}[0.6\tabcolsep]
                                                    &     $d$      &   110.5  &   -0.67  &   -81.5 &  -0.50  &  1.7   &   - &   -  \\
$^{23}$Na$|1, 1\rangle$ + $^{87}$Rb$|1, 1\rangle$  &     $s$       &   346.3  &   0.70 &   3.7 &  1.4  &  2.2     &  347.8   &  ~\cite{wang2013observation} \\
                                                   &     $p$       &   279.2  &   -2.4 &   20.1 &  1.5  &  2.2    &   284.1/284.2$^{\flat}$   &  ~\cite{wang2013observation}   \\
                                                    &     $p$      &   396.8 &   -1.2  &   19.8 &  1.0  &  1.7      &   -  &   - \\
                                                     &     $d$      &   268.1 &   -0.77  &   -30.1 &  -1.8  &  1.7   &   - &   -  \\
\rowcolor{mygray}[0.6\tabcolsep]
$^{23}$Na$|1, 1\rangle$ + $^{133}$Cs$|3, 3\rangle$  &     $d$     &   986.0   &   -1.4 &   14.3 &  3.8  &   2.4  &   - &   -  \\

\hline
\end{tabular}
\\
\begin{flushleft}

$\flat$ The observed doublet structure of a $p$-wave FR.
\end{flushleft}
\label{table3}
\end{table*}

\begin{table*}[h]
\renewcommand\arraystretch{1.4}
\caption{Same as Table~\ref{table1} but for the $^{39}$K-X, $^{40}$K-X and $^{41}$K-X mixtures, X being isotopes of K, Rb, or Cs atom. ``Broad" $s$-, $p$-, and $d$-wave FRs in the lowest-energy scattering channels are not found in the mixtures of $^{40}$K-$^{41}$K or $^{40}$K-$^{133}$Cs.}
\begin{tabular}{|p{3.9cm}|p{0.6cm}<{\centering}|p{1.3cm}<{\centering}|p{1.3cm}<{\centering}|p{1.3cm}<{\centering}|p{1.3cm}<{\centering}|p{1.3cm}<{\centering}|p{2.1cm}<{\centering}|p{2.7cm}<{\centering}|}
\hline
Scattering channel  &  $l$  &         $B_{0l}$ (G) &   $\zeta_{\textmd{res}}$    &   $\Delta_{Bl}$ (G)   &  $\tilde{a}_{\textmd{bgl}}/\bar{a}_{l} $   &   $\delta\mu_{l}/\mu_{B}$   &   $B_{0l}^{\textmd{expt}}$ (G) &  Remark/Reference  \\
\hline
$^{39}$K$|1, 1\rangle$ + $^{40}$K$|9/2, -9/2\rangle$  &     $s$    &   112.0  &   5.7      &   3.0 &  16.5  &   1.6    &   -   &   -  \\
                                                     &     $s$     &   133.6  &   94.0   &   -43.1 &  -  &   1.9      &   -   &   -   \\
                                                     &     $s$     &   143.6 &   1.3   &   -0.11  &  -76.4 &   1.7    &   -   &   -    \\
                                                     &     $s$     &   796.5 &   3.1   &   -0.33  &  -28.3 &   3.8    &   -   &   -    \\
\rowcolor{mygray}[0.6\tabcolsep]
$^{39}$K$|1, 1\rangle$ + $^{41}$K$|1, 1\rangle$    &     $p$   &   448.1   &   -0.66 &   -1.5 &  -4.3  &   2.0    &   -  &   -  \\
$^{39}$K$|1, 1\rangle$ + $^{85}$Rb$|2, 2\rangle$   &     $d$     &   363.7  &   -6.7    &   11.5 &  13.9  &   2.1     &   -   &   -  \\
                                                    &     $d$    &   464.9 &   -0.57    &   1.2 &  12.8  &  1.9        &   - &   -   \\
                                                   &     $d$     &   677.3 &   -0.51    &   0.43 &  16.2  &  3.7      &   -  &   -  \\
                                                    &     $d$    &   706.6 &   -2.8     &   2.7 &  14.8  &  3.6        &   -  &   -  \\
\rowcolor{mygray}[0.6\tabcolsep]
$^{39}$K$|1, 1\rangle$ + $^{87}$Rb$|1, 1\rangle$    &     $s$    &   317.2 &   1.3    &   7.6 &  0.50  &  2.5   &   317.9   &  ~\cite{Modugno2008KRb}   \\\rowcolor{mygray}[0.6\tabcolsep]
                                                   &     $p$    &   274.3  &   -0.69  &   -5.0 &  -0.67  &  2.5    &   277.57/277.70$^{\flat}$   &  ~\cite{Modugno2008KRb}     \\
\rowcolor{mygray}[0.6\tabcolsep]
                                                    &     $d$    &   186.0 &   -0.66  &   4.6 &  3.0  &  2.4    &   - &   -   \\
$^{39}$K$|1, 1\rangle$ + $^{133}$Cs$|3, 3\rangle$     &   $s$       &   361.0&  1.4       &   3.9 &  0.91  &  2.1     &   361.1   &  ~\cite{Hutson2017KCs}  \\
                                                      &     $s$     &   950.3&   0.64    &  1.1  &  0.93  &  3.5        &   - &   - \\
                                                      &     $p$     &   333.7&   -1.4    &  -33.9  &  -0.18  &  2.1     &   - &   - \\
                                                      &     $p$     &   919.8&   -0.66   &  -12.2  &  -0.14  &  3.5     &   -  &   - \\
                                                      &     $d$     &   254.6&   -19.2   &  31.7 &  11.1 &  2.1         &   -  &   - \\
                                                      &     $d$     &   359.0&   -0.98   &  2.4 &  8.3 &  1.9           &   - &   -  \\
                                                      &     $d$     &   812.0&   -1.4    &  0.80  &  18.7 &  3.6        &   - &   -  \\
                                                      &     $d$     &   848.6&   -15.7   &   12.2 &  14.1 &  3.5        &   - &   -  \\
\rowcolor{mygray}[0.6\tabcolsep]
$^{40}$K$|9/2, -9/2\rangle$ + $^{85}$Rb$|2, 2\rangle$  &     $s$      &   339.1 &   3.0     &   -35.6 &  -0.34  &  2.0     &   - &   -  \\\rowcolor{mygray}[0.6\tabcolsep]
                                                        &     $p$     &  288.5  &   -0.68   &   -3.1 &  -1.2  &  2.1        &   - &   -  \\
$^{40}$K$|9/2, -9/2\rangle$ + $^{87}$Rb$|1, 1\rangle$    &     $s$    &   547.7  &   1.7  &   -1.7 &  -2.9  &  2.4       &   546.89   &  ~\cite{Arlt2007KRb}  \\
                                                        &     $s$      &   628.4 &   0.87 &   -0.71 &  -3.0  &  2.8     &   659.68   &  ~\cite{Arlt2007KRb}     \\
\rowcolor{mygray}[0.6\tabcolsep]
$^{41}$K$|1, 1\rangle$ + $^{85}$Rb$|2, 2\rangle$   &     $s$   &   182.1   &   6.7   &   3.4 &  5.2  &  2.6  & - &   -   \\\rowcolor{mygray}[0.6\tabcolsep]
                                                   &     $s$    &  191.0   &  0.50   &   0.58 &  3.0  &  2.0  &   - &   -    \\\rowcolor{mygray}[0.6\tabcolsep]
                                                   &     $s$    &   656.1  &   9.6   &  4.1  &  8.8  &  1.9   &   -  &   -    \\\rowcolor{mygray}[0.6\tabcolsep]
                                                   &     $s$    &   680.7  &   19.0  &  17.1  &  4.4  &  1.8 &   -  &   -     \\\rowcolor{mygray}[0.6\tabcolsep]
                                                   &     $p$    &   668.2  &   -3.9  &  -8.8  &  -2.7  &  1.9     &   -  &   -      \\
$^{41}$K$|1, 1\rangle$ + $^{87}$Rb$|1, 1\rangle$   &     $s$      &   38.1   &   39.9     &   33.6 &  5.0  &  1.7       &   35.2   &  ~\cite{Inguscio2008K41Rb87}  \\
                                                     &     $s$    &   78.3   &   1.2      &  8.1  &  0.86  &  1.7         &   78.61   &  ~\cite{Inguscio2008K41Rb87}  \\
                                                   &     $s$      &   539.7  &   56.1     &  78.5  &  2.5  &  1.9       &   -&-  \\
                                                   &     $p$      &   121.9  &   -31.4    &  -41.9  &  -3.4  &  2.5     &   -  &- \\
                                                    &     $p$     &   850.8   &   -244.8  &  -220.4  &  -5.0  &  2.5     &   - &-  \\
                                                    &     $d$     &   41.8   &   -1.5     &  44.8  &  0.65  &  2.4       &   -  &-  \\
                                                    &     $d$     &   553.6  &   -4.0     &  230.5  &  0.40  &  2.2      &   -  &-  \\
\rowcolor{mygray}[0.6\tabcolsep]
$^{41}$K$|1, 1\rangle$ + $^{133}$Cs$|3, 3\rangle$   &     $s$    &   173.3 &  0.86     &   0.78 &  2.5  &  2.2       &   - &- \\\rowcolor{mygray}[0.6\tabcolsep]
                                                    &     $s$    &   911.0 &  3.1      &   3.0 &  2.5  &  2.1       &   - &-  \\\rowcolor{mygray}[0.6\tabcolsep]
                                                    &     $p$    &   105.7 &   -0.92   &  -0.40  &  -6.9  &  2.8    &   - &-  \\\rowcolor{mygray}[0.6\tabcolsep]
                                                    &     $p$    &   148.2 &   -4.7    &  -2.8  &  -6.8  &  2.1     &   - &-  \\\rowcolor{mygray}[0.6\tabcolsep]
                                                    &     $p$    &   154.8 &   -1.2    &  -0.53 &  -10.4 &  1.9     &   - &-  \\\rowcolor{mygray}[0.6\tabcolsep]
                                                    &     $p$    &   897.4 &   -18.4   &  -11.6 &  -6.3 &  2.1      &   - &-  \\\rowcolor{mygray}[0.6\tabcolsep]
                                                    &     $p$    &   960.6 &   -0.73   &  -0.42  &  -7.4  &  2.0     &   - &-  \\
\hline
\end{tabular}
\\
\begin{flushleft}
$\flat$ The observed doublet structure of a $p$-wave FR.
\end{flushleft}
\label{table4}
\end{table*}

\begin{table*}[h]
\renewcommand\arraystretch{1.5}
\caption{Same as Table~\ref{table1} but for the $^{85}$Rb-X and $^{87}$Rb-X mixtures, X being isotopes of Rb or Cs atom. }
\begin{tabular}{|p{3.6cm}|p{0.7cm}<{\centering}|p{1.3cm}<{\centering}|p{1.3cm}<{\centering}|p{1.3cm}<{\centering}|p{1.3cm}<{\centering}|p{1.3cm}<{\centering}|p{1.6cm}<{\centering}|p{2.7cm}<{\centering}|}
\hline
Scattering channel  &   $l$  &       $B_{0l}$ (G)  &   $\zeta_{\textmd{res}}$    &   $\Delta_{Bl}$ (G)   &  $\tilde{a}_{\textmd{bgl}}/\bar{a}_{l} $   &   $\delta\mu_{l}/\mu_{B}$   &   $B_{0l}^{\textmd{expt}}$ (G) &  Remark/Reference  \\
\hline
$^{85}$Rb$|2, 2\rangle$ + $^{87}$Rb$|1, 1\rangle$    &     $s$    &   530.4   &   84.7   &   70.1 &  2.6  &   1.6     &   569  &  ~\cite{You2016Rb85Rb87}    \\
                                                     &     $p$    &   668.3   &   -1.3   &   2.0 &  2.0 &   1.9       &   - &   - \\
                                                     &     $p$    &   813.0   &   -434.5 &   -200.2 &  -4.8 &   2.5   &   823.3 &  ~\cite{You2016Rb85Rb87}    \\
                                                     &     $d$    &   548.4   &   -6.7   &   213.9 &  0.42 &   1.9    &   622.6 &  ~\cite{You2017dwave}   \\
\rowcolor{mygray}[0.6\tabcolsep]
$^{85}$Rb$|2, 2\rangle$ + $^{133}$Cs$|3, 3\rangle$   &     $s$     &   112.2 &  24.2    &   240.7 &  0.21  &   1.8    &   107.13 &  ~\cite{Cornish2013RbCs}  \\\rowcolor{mygray}[0.6\tabcolsep]
                                                     &     $s$     &   187.9 &   1.1    &   -4.2 &  -0.39 &   1.7     &   187.66 &  ~\cite{Cornish2013RbCs}  \\\rowcolor{mygray}[0.6\tabcolsep]
                                                     &     $s$    &   631.9  &   29.6   &   264.8 &  0.10 &   3.1     &   641.8 &  ~\cite{Cornish2013RbCs}   \\\rowcolor{mygray}[0.6\tabcolsep]
                                                     &     $p$    &   72.3   &   -7.4   &   -17.3 &  -0.99 &   1.7    &   70.68 &  ~\cite{Cornish2013RbCs}   \\\rowcolor{mygray}[0.6\tabcolsep]
                                                      &     $p$    &   602.7  &   -9.2   &   -11.6 &  -0.99 &   3.1    &   614.6 &  ~\cite{Cornish2013RbCs}    \\\rowcolor{mygray}[0.6\tabcolsep]
                                                      &     $d$    &   6.3     &   -3.4     &   18.0 &  2.0 &   1.5  &  - &  -  \\\rowcolor{mygray}[0.6\tabcolsep]
                                                      &     $d$   &   555.1   &   -4.5  &   11.8 &  2.0 &   3.1    &   - &   -  \\
$^{87}$Rb$|1, 1\rangle$ + $^{133}$Cs$|3, 3\rangle$   &     $s$     &   287.0  &  4.5   & 0.54  &  6.7  &  2.9     &  279.03 &  ~\cite{Cornish2014Rb87Cs}   \\
                                                     &     $s$      &   310.8 &   5.5  &   0.74 &  6.4 &   2.7    &  310.72 &  ~\cite{Cornish2014Rb87Cs}   \\
                                                      &     $s$    &   337.5  &   2.8  &   0.41 &  6.2 &   2.5     &  352.7 &  ~\cite{Cornish2014Rb87Cs}    \\
                                                      &     $s$    &   834.7  &   2.5  &   0.49 &  6.3 &   1.9     &  790.2 &  ~\cite{Cornish2014Rb87Cs}    \\

\hline
\end{tabular}
\\
\label{table5}
\end{table*}

\FloatBarrier
\subsection{A unique $d$-wave shape resonance in $^{41}$K}\label{subsection3}

Among the resonances in all possible combinations of alkali-metal atoms, a unique $d$-wave shape resonance is identified in the lowest-energy scattering channel of $^{41}$K atoms, which was observed recently by Yao et al.~\cite{Yao2017dwave}. This is a particularly interesting resonance, not only because it has the advantage of being very ``broad" true single-channel resonance~\cite{Chin2010FRreview}, but also because it can be conveniently tuned into or out of resonant by applying a magnetic field like a normal magnetic FR.

\begin{figure}[h]
    \includegraphics[width=1\linewidth]{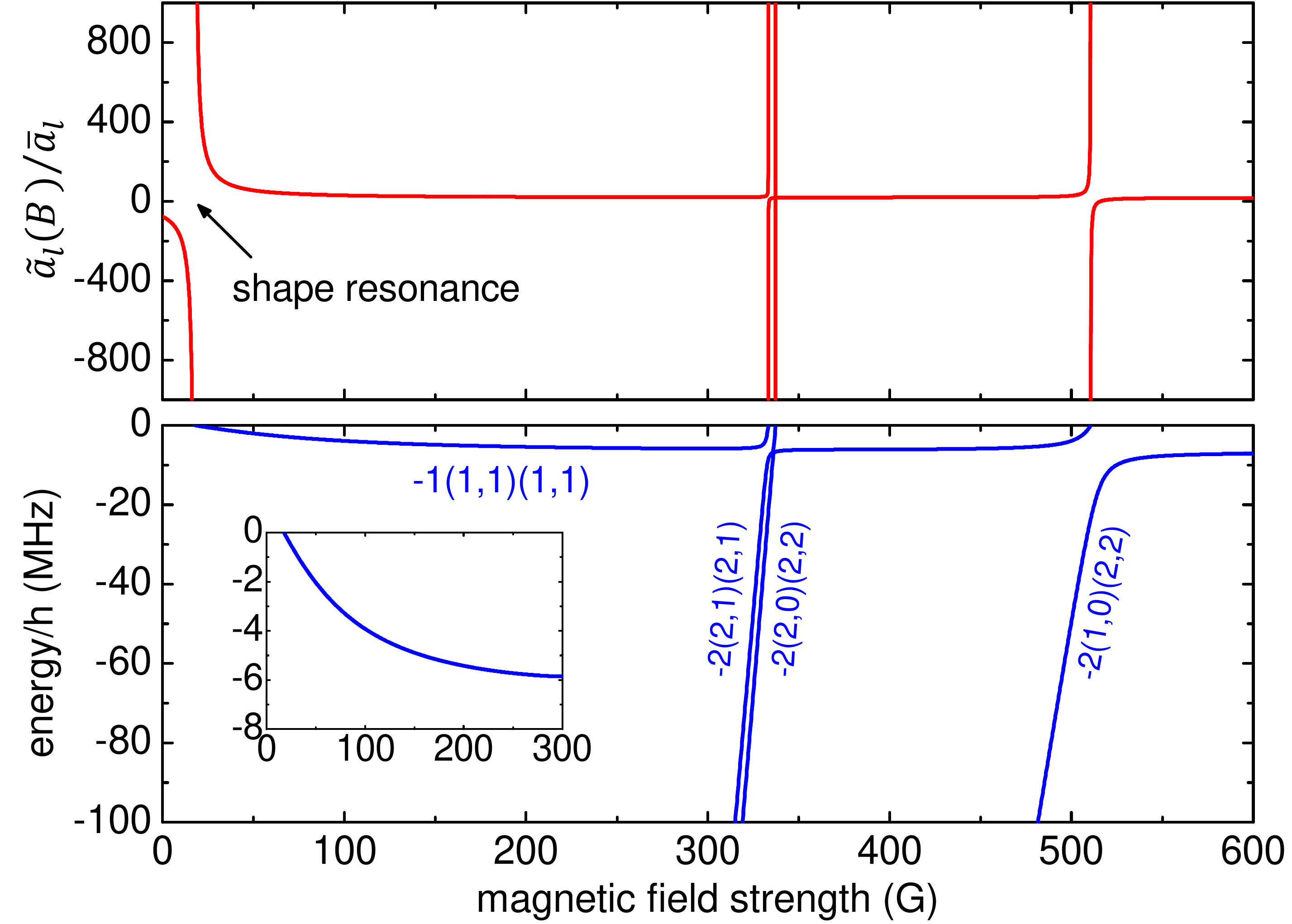}
    \caption{(Color online)
    Top panel: Calculated $d$-wave reduced generalized scattering length as a function of magnetic field for the $^{41}$K$|1,1\rangle$ + $^{41}$K$|1,1\rangle$ scattering channel. The black arrow denotes the position of the shape resonance. Bottom panel: Energies of the weakly bound molecular states relative to the dissociation threshold of the channel.  The bound states are labeled as $n$$(f_1,m_{f_1})$$(f_2,m_{f_2})$, where $n=-1,-2,...$ is the vibrational quantum number counting down from the top of the potential of the corresponding $^{41}$K$|f_1,m_{f_1}\rangle$ + $^{41}$K$|f_2,m_{f_2}\rangle$ channel. The inset shows an expanded view of the region near the position of the shape resonance.}
    \label{K41}
\end{figure}

The single-channel character of the resonance can be demonstrated by plotting the bound-state spectra of the scattering channel. The top panel of Fig.~\ref{K41} shows the dimensionless $d$-wave reduced generalized scattering length, $\tilde{a}_l(B)/\bar{a}_l$ (Eq.\,(\ref{al})), as a function of magnetic field for the $^{41}$K$|1,1\rangle$ + $^{41}$K$|1,1\rangle$ channel. The bottom panel shows the energies of the bound states relative to the dissociation threshold energy of the channel. The $d$-wave shape resonance mentioned above, which is located around 18\,G, occurs when the least bound $d$-wave state becomes degenerate with the threshold. The shape resonance comes with a width $|\Delta_{Bl}|$ of 146.5\,G and an effective $|\zeta_{\textmd{res}}|$ as large as 203.7. Besides the $d$-wave shape resonance, one ``broad" and two ``narrow" $d$-wave Feshbach resonances are also found in this scattering channel. The theoretical parameters of these resonances are listed in Table~\ref{table11}.

Even though shape resonances can offer very ``broad'' high partial wave interactions, they are typically not easily accessible due to the small magnetic sensitivity of the bound state energy with respect to the threshold energy in a single channel. The shape resonance in $^{41}$K represents a stroke of luck and is the only kind found in all alkali-metal systems, where the least bound $d$-wave state of the open channel is accidentally located very close to its threshold. For comparison, the $p$-wave shape resonance in $^{40}$K atoms and the $d$-wave one in $^{87}$Rb atoms can only be accessed by preparing the atoms with collision energies at $\sim k_B$(\SI{280}{\micro K})~\cite{Jin1999Kshaperesonance} ($k_B$ being the Boltzmann constant) and at $\sim k_B$(\SI{270}{\micro K})~\cite{Walraven2004Rbshaperesonance,Wilson2004Rbshaperesonance}, respectively, or by using indirect approaches~\cite{Kokkelmans2004Rbshaperesonance}. Therefore, the ``broad" shape resonance in $^{41}$K atoms mentioned above, which is easily accessible by tuning magnetic field, could find important applications in studying many-body physics with anisotropic interactions.

\begin{table}[h]
\renewcommand\arraystretch{1.5}
\caption{Calculated parameters for $d$-wave resonances in the lowest-energy scattering channel of $^{41}$K atoms, $^{41}$K$|1,1\rangle$ + $^{41}$K$|1,1\rangle$, for magnetic field in the range of 0-600\,G. The resonance located at 17.7\,G is a shape resonance and the others are Feshbach resonances.}

\begin{tabular}{p{1.6cm}<{\centering}p{1.6cm}<{\centering}p{1.6cm}<{\centering}p{1.6cm}<{\centering}p{1.6cm}<{\centering}}
\hline
\hline
 $B_{0l}$(G) &      $\zeta_{\textmd{res}}$   &  $\Delta_{Bl}$ (G)  &  $\tilde{a}_{\textmd{bgl}}/\bar{a}_{l} $   &   $\delta\mu_{l}/\mu_{B}$     \\
\hline
  17.7  &   -203.7    &   -146.5  &   10.0  &   -10.7   \\
 333.3  &    -0.42    &   0.42    &   20.5   &  3.8     \\
 337.3  &    -0.0072  &   0.0077  &  18.3  &   3.9       \\
  510.6  &   -2.6      &   5.4  &   19.0  &   2.0        \\
\hline
\hline
\end{tabular}
\\
\label{table11}
\end{table}

\section{Predicting and characterizing Feshbach resonances using MQDT}\label{section3}

In this study, a computationally simple MQDT assisted by the analytic wave functions for the long-range van-der-Waals (vdW) potential~\cite{Gao1998r6solutions,Gao2001AMQDT} is used to predict and describe FRs. This analytic MQDT approach has already been successfully applied and discussed in great length before~\cite{Bohn1998MQDT,Gao2005MQDT,Julienne2009threeparameter,Gao2011MQDTFR,Bohn2013highL,Gao2014LiKFR}. The main focus of this section is to provide its essential physical picture and to make our model and computations transparent for the less familiar readers.

The Hamiltonian that describes the collision between two alkali-metal atoms is given by
\begin{equation}\label{Hamiltonian}
\frac{\hbar^2}{2\mu}(-\frac{1}{r}\frac{d^2}{dr^2}r+\frac{\hat{l}^2}{r^2})+\hat{H}_1+\hat{H}_2+\hat{V}(r),
\end{equation}
where $\mu$ is the reduced mass, $r$ is the interatomic separation, $\hat{l}$ is the molecular orbital angular momentum operator, and $\hat{V}(r)$ is the interatomic-interaction potential operator. For alkali-metal atoms in the ground hyperfine manifold placed under a static magnetic field $B_z$, the free monomer Hamiltonians $\hat{H}_1$ and $\hat{H}_2$ are given by
\begin{equation}\label{freeatomH}
\hat{H}_i=\zeta_i\hat{i}_i\cdot\hat{s}_i+(g_e\mu_B\hat{s}_{i_z}+g_n\mu_B\hat{i}_{i_z})B_z,
\end{equation}
where $\zeta_i$ denotes the hyperfine coupling constant, $\hat{i}_i$ and $\hat{s}_i$ are the nuclear and electronic spin operators of atom $i$ ($i=1,2$), $\mu_B$ is the Bohr magneton, $g_n$ and $g_e$ are the nuclear and electronic $g$-factors. Eigenstates of the Hamiltonian in Eq.\,(\ref{freeatomH}) define the asymptotic magnetic-field dressed hyperfine basis $|f_{1}m_{f_1}\rangle|f_{2}m_{f_2}\rangle$ used in our calculation.

In our model, the interaction term $\hat{V}(r)$ includes only the dominating isotropic electronic Born-Oppenheimer (BO) potentials~\cite{Chin2010FRreview}. The weak anisotropic spin-dependent interactions, which include the magnetic dipole-dipole~\cite{Verhaar1988dipole,Axelsson1995dipole}
and the second order spin-orbit \cite{Krauss1996spinorbit,Julienne2000spinorbit} interactions, are ignored. Dropping the weaker interactions is not too much of a concern for our search of ``broad" FRs, since the FRs induced by the weak anisotropic coupling are expected to be ``narrow". Besides, doing so also removes the coupling between open and closed channels with different partial wave $l$ as well as those with different $m_{f_1}+m_{f_2}$, thereby greatly simplifies the computations. 

An $N$-channel scattering problem can generally be described by a set of $N$ linearly independent wave functions
\begin{equation}\label{wavefunction}
\psi_j=\sum_{i=1}^N\phi_{i}(\tau)F_{ij}(r)/r, \quad j=1,2,...,N.
\end{equation}
Here, $\phi_{i}(\tau)$ denotes the function of channel $i$ describing all degrees of freedom except the interatomic separation $r$. Constructing an $N\times N$ matrix $\mathbf{F}(r)$ with elements $F_{ij}(r)$, one can show that $\mathbf{F}(r)$ satisfies the close-coupled radial Schr\"odinger equation
\begin{equation}\label{CCequation}
\frac{d^{2}\mathbf{F}}{dr^{2}}+\frac{2\mu}{\hbar^{2}}[E\mathbf{I}-\mathbf{W}(r)]\mathbf{F}(r)=0,
\end{equation}
where $E$ is the total energy, $\mathbf{I}$ is the identity matrix, and $\mathbf{W}(r)$ denotes the coupling matrix with elements
\begin{equation}\label{Wr1}
W_{ji}(r)\!=\!\int\!\phi_{j}^*(\tau)\!\left[\frac{\hbar^2}{2\mu}\frac{\hat{l}^2}{r^2}\!+\hat{H}_1\!+\hat{H}_2\!+\hat{V}(r)\right]\!\phi_{i}(\tau)d\tau.
\end{equation}

In principle, with the complete knowledge of $W_{ji}(r)$, it is possible to predict FR positions with very high accuracy (within 1-Gauss uncertainty) by solving Eq.~(\ref{CCequation}) numerically. In practice, however, determining $W_{ji}(r)$ requires Herculean effort in data-fitting $W_{ji}(r)$ to vast amount of experimentally measured spectroscopic values, an overwhelming process consuming more time than in the current work (cf. \cite{Tiemann2010Rbpotential}).

The MQDT we adopt~\cite{Fano1982QDT,Seaton1983QDT,Kurti1988QDT,Bohn1998MQDT,Mies2000MQDT,Gao2005MQDT,Julienne2011MQDT} greatly simplifies the task to predict FRs. It takes advantage of the large differences in the energy and length scales between the short-range and the long-range potentials in $\hat V(r)$. As $r$ is reduced below the exchange-interaction range $r_0$ (typically around 30 a.u.), the energy splitting between the singlet- and triplet-state potentials in $\hat V(r)$ gradually overtakes the hyperfine and Zeeman energies in $\hat H_i$. Since the maximum depths of the singlet and triplet potentials are of the order of 10 to 100 THz, the long-range dominating $\hat H_i$ which are of the magnitudes of a few GHz have essentially no influence on the short-range ($r<r_0$) wave functions. Therefore, for the magnetic field range of 0-1000\,G considered in this work, one can effectively use two constants, $K_s^c$ and $K_t^c$, to reflect the overall effects of the singlet and triplet potentials to the long-range wave functions, negating the need for precise knowledges of the short-range part of $\mathbf{W}(r)$. Physically, the two quantum defects define the boundary conditions at $r_0$ for the long-range wave functions.

For $r\geq r_0$, the exchange interaction can be neglected and $\mathbf{W}(r)$ becomes diagonal in the asymptotic channel basis (namely, the eigen-energy basis of $\hat H_1+\hat H_2$), giving
\begin{equation}\label{Wr2}
W_{ij}(r)\stackrel{r\geq r_0}{\longrightarrow}\left[E_{i}^{\infty}+\frac{\hbar^{2}l_i(l_i+1)}{2\mu r^2}-\frac{C_6}{r^6}\right]\delta_{ij},
\end{equation}
where $E_i^{\infty}$ and $l_i$ are, respectively, the threshold energy and orbital angular-momentum quantum number for channel $i$, and $-{C_6}/{r^6}$ is the long-range vdW potential. The higher-order dispersion terms (such as $-C_8/r^8$ and $-C_{10}/r^{10}$) in the long-range potential are ignored. While such omission reduces the accuracy of our predictions, it allows us to make use of the analytic wave functions for the vdW potential, thereby enables highly efficient numerical computations.

The Schr\"odinger equation with the potential of Eq.\,(\ref{Wr2}) is solved analytically in Ref.~\cite{Gao1998r6solutions}, and a pair of linearly independent base functions $f_{\epsilon_sl}^{c}(r_s)$ and $g_{\epsilon_sl}^{c}(r_s)$ that are insensitive to energy and partial-wave variations at short-range ($r_s\to0$) can be defined [as given by Eq.~(12) of Ref.~\cite{BoGao2004vdW}]. Here, $r_s=r/\beta_6$ with $\beta_{6}=(2\mu{C_6}/{\hbar^2})^{1/4}$ being the length scale of the vdW interaction. Employing the MQDT as discussed earlier, the long-range ($r>r_0$) solutions of Eq.\,(\ref{CCequation}) can be written as
\begin{equation}\label{Fr}
\mathbf{F}(r)=\left[\mathbf{f}(r_s)-\mathbf{g}(r_s)\mathbf{K}^{c}\right]\mathbf{A}.
\end{equation}
Here, $\mathbf{f}(r_s)$ and $\mathbf{g}(r_s)$ are $N\times N$ diagonal matrices with elements $f_{ii}(r_s)\equiv f_{\epsilon_{si}l_i}^{c}(r_s)$ and $g_{ii}(r_s)\equiv g_{\epsilon_{si}l_i}^{c}(r_s)$, respectively. $\epsilon_{si}=(E-E_i^\infty)/s_E$  with $s_E=\hbar^2/(2\mu\beta^2_6)$ being the energy scale of the vdW interaction. The elements of the $N\times N$ matrix $\mathbf{K}^{c}$ are completely determined by the two short-range parameters $K_s^c$ and $K_t^c$ (Appendix~\ref{appendixA}), and $\mathbf{f}(r_s)-\mathbf{g}(r_s)\mathbf{K}^{c}$ represents a set of solutions that satisfy the short-range boundary conditions. The matrix $\mathbf{A}$, which is to be determined by the long-range boundary conditions at $r\to\infty$, ensures that the final solutions are physical.

The scattering problem considered in this work has only one open channel and $N_c=N-1$ closed channels, i.e. $E^\infty_1<E$ and $E^\infty_{i>1}>E$. The long-range boundary condition requires the closed-channel components to vanish at $r\to\infty$, namely, $\lim\limits_{r\to\infty}F_{ij}(r)\rightarrow0$ for $i>1$. Under this condition, there is only one physical solution and its long-range wave function is given by
\begin{equation}
\psi_1(r>r_0)= \frac{\phi_1(\tau)}{r}[f_{\epsilon_{s1}l_1}^{c}(r_s) + g_{\epsilon_{s1}l_1}^{c}(r_s)K^c_{\rm eff}(\epsilon_{s1},B)].
\end{equation}
Here,
\begin{equation}\label{Keff}
K^c_{\rm eff}(\epsilon_{s1},B)=K^c_{\rm oo}+K^c_{\rm oc}(\chi^c-K^c_{\rm cc})^{-1}K^c_{\rm co},
\end{equation}
where $\chi^c$ is a $N_c\times N_c$ diagonal matrix with elements
$\chi^c_{ii}\equiv\lim\limits_{r_s\to\infty}\frac{f_{\epsilon_{si}l_i}^{c}(r_s)}{g_{\epsilon_{si}l_i}^{c}(r_s)}$ given by Eq.\,(54) in Ref.~\cite{Gao2009singlechannel}, $K^c_{\rm oo}$, $K^c_{\rm oc}$, $K^c_{\rm co}$, and $K^c_{\rm cc}$ (with subscripts $``\rm o"$ and $``\rm c"$ referring to open and closed channels) are sub-matrices of $\mathbf{K}^c \equiv
\begin{bmatrix}
    K^c_{\rm oo} & K^c_{\rm oc}  \\
    K^c_{\rm co} & K^c_{\rm cc}
\end{bmatrix}
$.

In calculating FRs of atoms at ultracold temperatures, we set $\epsilon_{s1}=0$ and will denote $K^c_{\rm eff}(0,B)$ simply as $K^c_{\rm eff}(B)$ hereafter. The generalized scattering length for the $l$-th partial wave is related to $K^c_{\rm eff}(B)$ by~\cite{Gao2011MQDTFR}
\begin{equation}\label{al}
\tilde{a}_{l}(B)\!=\!\bar{a}_{l}\left[(-1)^l+\frac{1+\tan[(2l+1)\pi/8]K^c_{\rm eff}(B)}{K^c_{\rm eff}(B)-\tan[(2l+1)\pi/8]}\right],
\end{equation}
where $\bar{a}_{l}$ = $\bar{a}_{sl}$$\beta_{6}^{2l+1}$ is the mean scattering length for the $l$-th partial wave, $\bar{a}_{sl}$ being an $l$-dependent constant~\cite{aslformula}. A FR occurs when $\tilde{a}_{l}(B)$ diverges to infinity, i.e. when the denominator of Eq.~(\ref{al}) becomes zero:
\begin{equation}\label{B0l}
K^c_{\rm eff}(B_{0l})-\tan[(2l+1)\pi/8]=0.
\end{equation}
Around the resonance position $B_{0l}$, $\tilde{a}_{l}(B)$ varies approximately as $\tilde{a}_{l}(B) = \tilde{a}_{\textmd{bgl}}[1 - \Delta_{Bl}/(B-B_{0l})]$~\cite{Chin2010FRreview}, a form which defines the parameters we tabulate for each FR. Using these parameters, the resonance strength $\zeta_{\textmd{res}}$ can be obtained by~\cite{Gao2011MQDTFR}
\begin{equation}\label{zeta}
\zeta_{\textmd{res}} = -\frac{1}{(2l+3)(2l-1)} \frac{\tilde{a}_{\textmd{bgl}}}{\bar{a}_{l}} \left( \frac{\delta\mu_{l} \Delta_{Bl}}{s_{E}} \right).
\end{equation}
The parameter $\zeta_{\textmd{res}}$ is used to define the ``broadness'' of a FR. The differential magnetic moment $\delta\mu_{l}$ is defined by $\delta\mu_{l}=d\epsilon_l/dB|_{B=B_{0l}}$, where the molecular state energy $\epsilon_l$ is obtained by solving~\cite{Gao2011MQDTFR}
\begin{equation}\label{elB}
K^c_{\rm eff}(\epsilon_l,B)-\tan[(2l+1)\pi/8]=0.
\end{equation}

Within the MQDT model discussed above, the prediction of FRs in arbitrary partial waves requires just three parameters ~\cite{Gao2005MQDT,Julienne2009threeparameter,Gao2014LiKFR}: the vdW coefficient $C_6$ and the singlet (triplet) scattering length $a_s$ ($a_t$) which is related to $K_s^c$ ($K_t^c$) (Appendix\,\ref{appendixB}), in addition to parameters for inherent atomic properties such as the atomic masses and hyperfine splittings. The predictive power of the analytic MQDT has been proven previously in systems such as $^{40}$K-$^{87}$Rb~\cite{Julienne2009threeparameter}, $^{6}$Li-$^{40}$K~\cite{Gao2014LiKFR}, and $^{85}$Rb-$^{87}$Rb mixtures~\cite{You2016Rb85Rb87,You2017dwave}. The predictions by our MQDT are generally not as accurate as those offered by numerical coupled-channel calculations based on the full knowledge of the molecular potentials of the collision pairs. However, at the expense of accuracy, the simplification made by MQDT gains substantial advantages in efficiency and in negating the need of short-range molecular potentials, the latter is particular useful since not all the molecular potentials of alkali-metal systems are well known.

In our investigation of FRs in alkali-metal systems, we optimize the three parameters $C_6$, $a_s$ and $a_t$ to minimize the discrepancies between the calculated and the measured positions of previously known resonances. However, for systems without available experimental data, the accuracy of the predictions is then determined by the reliability of the three input parameters taken from the relevant references and the applicability of the simplified MQDT model. The parameters we adopt for all the alkali-metal systems are listed in Appendix~\ref{appendixB}. We caution that these numbers are not to be taken as more accurate than those adopted in other coupled channel calculations.

\section{Conclusion}\label{section4}
In summary, we report the calculated results for all the ``broad" $s$-, $p$-, and $d$-wave FRs free from two-body loss in alkali-metal systems for magnetic field range of 0-1000\,G, based on a simple analytic MQDT. A number of systems exhibiting ``broad" high partial-wave FRs and several extraordinarily ``broad" resonances are identified and highlighted. Our results help to categorize systems suitable for experimental studies of universal properties in strongly interacting atomic gases. We find that ``broad" $p$- or $d$-wave FRs in the lowest-energy scattering channels do not exist for all possible fermionic alkali-metal combinations. This encourages further explorations for low-loss ``broad" high partial-wave FRs in fermionic atoms to go beyond alkali-metal systems.

Once deciding on a system or a resonance of interest for further experimental study, a more precise characterization is possible if needed. For a single resonance, this can be achieved by measuring the binding energies of the corresponding Feshbach molecule and following the analysis embedded in Ref~\cite{Gao2011MQDTFR}. At a system level, including all its resonances, improved characterization can be achieved using CC calculations (see, e.g., \cite{Tiemann2010Rbpotential}) or numerical MQDT (see, e.g., ~\cite{Bohn2013highL}) if precise potentials are available, or using multiscale MQDT along the line of Ref.~\cite{Fu16a} to take into account potentials of shorter range, such as the $-C_8/r^8$ potential.

\begin{acknowledgments}
We thank Gaoren Wang for helpful discussions. This work is supported at Tsinghua by the National Key R$\&$D Program of China (Grants No. 2018YFA0306503 and 2018YFA0306504) and by NSFC (Grants No. 91636213, No. 11574177, No. 91736311), and at Toledo by NSF (Grant No. PHY-1607256).
\end{acknowledgments}

\appendix
\section{Obtaining $\mathbf{K}^c$ from the short-range parameters $K^c_{s}$ and $K^c_{t}$}\label{appendixA}

The basis that diagonalizes the short-range ($r<r_0$) BO potentials is the $|IM_{I}SM_{S}\rangle$ basis characterized by quantum numbers of the total nuclear spin $\mathbf{I}=\mathbf{i}_1+\mathbf{i}_2$ and total electronic spin $\mathbf{S}=\mathbf{s}_1+\mathbf{s}_2$. Furthermore, since the functions $f_{\epsilon_sl}^{c}(r_s)$ and $g_{\epsilon_sl}^{c}(r_s)$ are essentially independent of $\epsilon_s$ and $l$ at short-range (where $|\hat V(r)|\gg|E-E_i^\infty|$), one can approximate them by, say, $\bar {f^c}(r_s)$ and $\bar{g^c}(r_s)$, respectively, around $r_0$. Together, the two aforementioned properties mean that, near $r_0$, the spatial wave function in the $|IM_{I}SM_{S}\rangle$ basis can be most conveniently written as
\begin{equation}\label{FrIS}
\mathbf{F}^{(IS)}(r)\approx[\bar{f^c}(r_s)\mathbf{I}-\bar{g^c}(r_s)\mathbf{I}\mathbf{K}^{c(IS)}]\mathbf{A}^{(IS)},
\end{equation}
where $\mathbf{K}^{c(IS)}$ is a diagonal matrix with elements
\begin{widetext}
\begin{equation}\label{KcIS}
\langle IM_{I}SM_{S}|\mathbf{K}^{c(IS)}|I'M'_{I}S'M'_{S}\rangle=\delta_{II'}\delta_{M_{I}M'_{I}}\delta_{SS'}\delta_{M_{S}M'_{S}}K^{c(IS)}_{S,M_S}.
\end{equation}
Here, $K^{c(IS)}_{S,M_S}$ depends only on whether the channel is an electronic singlet or triplet state, namely, $K^{c(IS)}_{0,0}=K^c_{s}$ and $K^{c(IS)}_{1,1}=K^{c(IS)}_{1,0}=K^{c(IS)}_{1,-1}=K^c_{t}$. Comparing Eq.~(\ref{FrIS}) to Eq.~(\ref{Fr}) under the basis of $|f_{1}m_{f_1}\rangle|f_{2}m_{f_2}\rangle$ and noting that $\mathbf{F}(r)=U\mathbf{F}^{(IS)}(r)$ for a given $l$, one can show that $\mathbf{K}^c = U\mathbf{K}^{c(IS)}U^\dag$, where $U$ is a unitary matrix which transforms the basis $|IM_{I}SM_{S}\rangle$ to $|f_{1}m_{f_1}\rangle|f_{2}m_{f_2}\rangle$. Based on this relation, one can obtain the elements of $\mathbf{K}^c$ as shown below. (Note that in practice, one only needs to include channels with the same
  $M_F$ and $l$ as those of the open channel when computing $\mathbf{K}^c_{ij}$ using Eqs.~(\ref{Kij}) and (\ref{Kexij}).)

\subsection{Heteronuclear system}\label{subsection4}
The matrix elements $K^c_{ij}$ are given by
\begin{eqnarray}\label{KcHF}
K^c_{ij}&&=\langle f_{2}^im_{f_2}^i|\langle f_{1}^im_{f_1}^i|\mathbf{K}^{c}|f_{1}^jm_{f_1}^j\rangle|f_{2}^jm_{f_2}^j\rangle \nonumber\\
&&=\sum_{IM_{I}SM_{S}I'M'_{I}S'M'_{S}}\langle f_{2}^im_{f_2}^i|\langle f_{1}^im_{f_1}^i|IM_{I}SM_{S}\rangle\langle IM_{I}SM_{S}|\mathbf{K}^{c(IS)}|I'M'_{I}S'M'_{S}\rangle\langle I'M'_{I}S'M'_{S}|f_{1}^jm_{f_1}^j\rangle|f_{2}^jm_{f_2}^j\rangle \nonumber\\
&&=\sum_{IM_{I}SM_{S}I'M'_{I}S'M'_{S}}\langle f_{2}^im_{f_2}^i|\langle f_{1}^im_{f_1}^i|IM_{I}SM_{S}\rangle\delta_{II'}\delta_{M_{I}M'_{I}}\delta_{SS'}\delta_{M_{S}M'_{S}}K^{c(IS)}_{S,M_S}\langle I'M'_{I}S'M'_{S}|f_{1}^jm_{f_1}^j\rangle|f_{2}^jm_{f_2}^j\rangle \nonumber\\
&&=\sum_{IM_ISM_{S}}\langle f_{2}^im_{f_2}^i|\langle f_{1}^im_{f_1}^i|IM_ISM_{S}\rangle\langle IM_ISM_{S}|f_{1}^jm_{f_1}^j\rangle|f_{2}^jm_{f_2}^j\rangle K^{c(IS)}_{S,M_S} \nonumber\\
&&=\sum_{SM_{S}}\omega_{ij}^{S,M_S}K^{c(IS)}_{S,M_S},
\end{eqnarray}
with $\omega_{ij}^{S,M_S}=\sum\limits_{IM_I}\langle f_{2}^im_{f_2}^i|\langle f_{1}^im_{f_1}^i|IM_ISM_{S}\rangle\langle IM_ISM_{S}|f_{1}^jm_{f_1}^j\rangle|f_{2}^jm_{f_2}^j\rangle$. The $\omega_{ij}^{S,M_S}$ can be expanded as
\begin{eqnarray}\label{omega}
\omega_{ij}^{S,M_S}&=&\sum_{IM_I}~~\sum_{i_1m_{i_1}s_1m_{s_1}i_2m_{i_2}s_2m_{s_2}}\langle f_{2}^im_{f_2}^i|\langle f_{1}^im_{f_1}^i|i_1m_{i_1}s_1m_{s_1},i_2m_{i_2}s_2m_{s_2}\rangle\langle i_1m_{i_1}s_1m_{s_1},i_2m_{i_2}s_2m_{s_2}|IM_ISM_{S}\rangle \nonumber\\ &&\sum_{i'_1m'_{i_1}s'_1m'_{s_1}i'_2m'_{i_2}s'_2m'_{s_2}}\langle IM_ISM_{S}|i'_1m'_{i_1}s'_1m'_{s_1},i'_2m'_{i_2}s'_2m'_{s_2}\rangle\langle i'_1m'_{i_1}s'_1m'_{s_1},i'_2m'_{i_2}s'_2m'_{s_2}|f_{1}^jm_{f_1}^j\rangle|f_{2}^jm_{f_2}^j\rangle.
\end{eqnarray}
In calculating Eq.~(\ref{omega}), the $|f_{k}m_{f_k}\rangle$ states of the two atoms ($k=1,2$) for channel $i$ or $j$ need to be expanded as follow
\begin{eqnarray}\label{ISHF}
|f_{1}^{i(j)}m_{f_1}^{i(j)}\rangle=a_1^{i(j)}|m_{s_{1}}=\frac{1}{2},m_{i_{1}}=m_{f_{1}}^{i(j)}-\frac{1}{2}\rangle+b_1^{i(j)}|m_{s_{1}}=-\frac{1}{2},m_{i_{1}}=m_{f_{1}}^{i(j)}+\frac{1}{2}\rangle \nonumber\\
|f_{2}^{i(j)}m_{f_2}^{i(j)}\rangle=a_2^{i(j)}|m_{s_{2}}=\frac{1}{2},m_{i_{2}}=m_{f_{2}}^{i(j)}-\frac{1}{2}\rangle+b_2^{i(j)}|m_{s_{2}}=-\frac{1}{2},m_{i_{2}}=m_{f_{2}}^{i(j)}+\frac{1}{2}\rangle,
\label{dressedhyperfinestate}
\end{eqnarray}
where the $|m_{s_k},m_{i_k}\rangle$ is the simplified notation of the uncoupled $|s_km_{s_k},i_km_{i_k}\rangle$ basis for atom $k$ ($k=1,2$).
The expansion coefficients $a$ and $b$ can be easily obtained from the Breit-Rabi formula~\cite{Gao2014LiKFR,Rabi1931BR}.

Making use of Eqs.\,(\ref{KcHF}-\ref{ISHF}) and the properties of the Clebsch-Gordan coefficients, one can show that $K^c_{ij}$ is given by
\renewcommand\arraystretch{1.5}
\begin{equation}\label{Kij}
K^c_{ij} = \left\{
             \begin{array}{lcl}
             {D_{ij}K^c_{s}+(1-D_{ij})K^c_{t}},\quad  &&i=j, \\
             {D_{ij}(K^c_{s}-K^c_{t})},\quad  &&i\neq j,
             \end{array}
        \right.
\end{equation}
where
\renewcommand\arraystretch{1.5}
\begin{equation}\label{Dij}
D_{ij} = \left\{
             \begin{array}{lcl}
             {\frac{1}{2}(a_1^ib_2^ia_1^jb_2^j+b_1^ia_2^ib_1^ja_2^j)},\quad  &&\Delta m_{f_1}=0, \\
             {-\frac{1}{2}a_1^ib_2^ib_1^ja_2^j},\quad &&\Delta m_{f_1}=-1, \\
             {-\frac{1}{2}b_1^ia_2^ia_1^jb_2^j},\quad &&\Delta m_{f_1}=+1, \\
             {0},\quad &&\mathrm{else}, \\
             \end{array}
        \right.
\end{equation}
with $\Delta m_{f_1}$ defined as $\Delta m_{f_1}=m_{f_1}^j-m_{f_1}^i$. Due to the conservation of $m_{f_1}+m_{f_2}$, $\Delta m_{f_2}=m_{f_2}^j-m_{f_2}^i=-\Delta m_{f_1}$.

\subsection{Homonuclear system}\label{subsection5}
For homonuclear systems with indistinguishable atoms, the total wave function of the systems should satisfy the permutation symmetries of bosonic or fermionic particles. If $|f_{1}m_{f_1}\rangle$ and $|f_{2}m_{f_2}\rangle$ are not the same, the spin part of the wave function is written as
\begin{equation}\label{homo1}
\frac{1}{\sqrt{2}}(|f_{1}m_{f_1}\rangle|f_{2}m_{f_2}\rangle+(-1)^{(2I+1+l)}|f_{2}m_{f_2}\rangle|f_{1}m_{f_1}\rangle),
\end{equation}
where $2I+1$ is even for bosonic and odd for fermionic atoms. If $|f_{1}m_{f_1}\rangle$ and $|f_{2}m_{f_2}\rangle$ are the same, the spin wave function is written as
\begin{equation}\label{homo2}
\frac{1}{2}(|f_{1}m_{f_1}\rangle|f_{1}m_{f_1}\rangle+(-1)^{(2I+1+l)}|f_{1}m_{f_1}\rangle|f_{1}m_{f_1}\rangle),
\end{equation}
which is different from Eq.~(\ref{homo1}) only in the normalization constant. Eq.~(\ref{homo2}) implies that for collisions of identical atoms in the same hyperfine state, only even $l$ is allowed for bosons and only odd $l$ for fermions. The matrix elements of $\mathbf{K}^c$ for homonuclear systems are therefore given by
\begin{eqnarray}\label{Kchomo}
K^c_{\mathrm{homo},ij}
&&=C_i(\langle f_{2}^im_{f_2}^i|\langle f_{1}^im_{f_1}^i|\pm\langle f_{1}^im_{f_1}^i|\langle f_{2}^im_{f_2}^i|)\mathbf{K}^{c}C_j(|f_{1}^jm_{f_1}^j\rangle|f_{2}^jm_{f_2}^j\rangle\pm|f_{2}^jm_{f_2}^j\rangle|f_{1}^jm_{f_1}^j\rangle) \nonumber\\
&&=C_iC_j\langle f_{2}^im_{f_2}^i|\langle f_{1}^im_{f_1}^i|\mathbf{K}^{c}|f_{1}^jm_{f_1}^j\rangle|f_{2}^jm_{f_2}^j\rangle\pm C_iC_j\langle f_{2}^im_{f_2}^i|\langle f_{1}^im_{f_1}^i|\mathbf{K}^{c}|f_{2}^jm_{f_2}^j\rangle|f_{1}^jm_{f_1}^j\rangle \nonumber\\
&&\quad\pm C_iC_j\langle f_{1}^im_{f_1}^i|\langle f_{2}^im_{f_2}^i|\mathbf{K}^{c}|f_{1}^jm_{f_1}^j\rangle|f_{2}^jm_{f_2}^j\rangle+C_iC_j\langle f_{1}^im_{f_1}^i|\langle f_{2}^im_{f_2}^i|\mathbf{K}^{c}|f_{2}^jm_{f_2}^j\rangle|f_{1}^jm_{f_1}^j\rangle,
\end{eqnarray}
where the $\pm$ sign is determined by the parity of $2I+1+l$.
\end{widetext}
The normalization constant $C_i$ and $C_j$ are
\renewcommand\arraystretch{1.5}
\begin{equation}\label{Ci}
C_{i(j)} = \left\{
             \begin{array}{lcl}
             {1/2}, \quad  &&f_{1}^{i(j)}=f_{2}^{i(j)},m_{f_{1}}^{i(j)}=m_{f_{2}}^{i(j)}, \\
             {1/\sqrt{2}},\quad &&\mathrm{else}. \\
             \end{array}
        \right.
\end{equation}
The first and the fourth term in Eq.~(\ref{Kchomo}) are the same as $K^c_{ij}$ in the heteronuclear case except for a coefficient $C_iC_j$. Denoting the second and third exchange terms as $K^c_{\mathrm{ex},ij}$, Eq.~(\ref{Kchomo}) can be written as
\begin{equation}\label{Kchomo2}
K^c_{\mathrm{homo},ij}=2C_iC_j\left[K^c_{ij}+(-1)^{(2I+1+l)}K^c_{\mathrm{ex},ij}\right],
\end{equation}
with
\renewcommand\arraystretch{1.5}
\begin{equation}\label{Kexij}
K^c_{\mathrm{ex},ij} = \left\{
             \begin{array}{lcl}
             {E_{ij}K^c_{s}+(1-E_{ij})K^c_{t}},\quad  &&i=j, \\
             {E_{ij}(K^c_{s}-K^c_{t})},\quad  &&i\neq j,
             \end{array}
        \right.
\end{equation}
where
\renewcommand\arraystretch{1.5}
\begin{equation}\label{Eij}
E_{ij} = \left\{
             \begin{array}{lcl}
             {\frac{1}{2}(a_1^ib_2^ia_2^jb_1^j+b_1^ia_2^ib_2^ja_1^j)},\quad  &&\Delta m_{f_1}=0, \\
             {-\frac{1}{2}a_1^ib_2^ib_2^ja_1^j},\quad &&\Delta m_{f_1}=-1, \\
             {-\frac{1}{2}b_1^ia_2^ia_2^jb_1^j},\quad &&\Delta m_{f_1}=+1, \\
             {0},\quad &&\mathrm{else}. \\
             \end{array}
        \right.
\end{equation}
\section{Input parameters $a_s$, $a_t$, and $C_{6}$}\label{appendixB}

The singlet scattering length $a_s$, the triplet scattering length $a_t$, and the vdW coefficient $C_{6}$ adopted in this work for every alkali-metal system are listed in Table~\ref{table6}. For systems with available experimental data on FRs, these parameters are optimized to give the best overall agreements between the predictions based on our theory and the experimentally measured resonance positions. As such, some of them are adjusted from the values given in the references.
The singlet and triplet scattering lengths $a_s$ and $a_t$  are related to $K_{s}^c$ and $K_{t}^c$ according to~\cite{Gao2005MQDT}
\begin{equation}\label{Kc}
a_{s(t)}/\beta_6=\left[\frac{\Gamma(3/4)}{2\Gamma(5/4)}\right]\frac{K_{s(t)}^c+\tan(\pi/8)}{K_{s(t)}^c-\tan(\pi/8)}.
\end{equation}

\begin{table}[H]
\renewcommand\arraystretch{1.5}
\caption{The optimized parameters $a_s$, $a_t$ and $C_{6}$ for calculating FRs in alkali-metal systems by analytic MQDT.}
\begin{tabular}{p{1.6cm}p{1.6cm}<{\centering}p{1.6cm}<{\centering}p{1.5cm}<{\centering}p{1.6cm}<{\centering}}
\hline
\hline
 System   &    $a_s$ (a.u.)    &   $a_t$ (a.u.)   &   $C_{6}$ (a.u.)     &  Reference    \\
\hline
$^{6}$Li-$^{6}$Li  & 44.45${^\flat}$   &    -2040${^\flat}$   &   1393.39  &   ~\cite{Hulet1997Li,Yan1996LiC6}  \\
$^{6}$Li-$^{7}$Li   &  -20  &   40.9   &   1390  &   ~\cite{Kokkelmans2004Li6Li7,Hulet1997Li}  \\
$^{6}$Li-$^{23}$Na   &  21${^\flat}$  &   14${^\flat}$   &   1467  &   ~\cite{Robin2008LiNa,Dalgarno2001C6}  \\
$^{6}$Li-$^{39}$K   &  64.4  &    67.7   &   2322  &   ~\cite{Tieckethessis}  \\
$^{6}$Li-$^{40}$K   &  52.5  &    63.7  &   2322  &   ~\cite{Gao2014LiKFR}  \\
$^{6}$Li-$^{41}$K   &  42.3  &    60.38${^\flat}$   &   2322  &   ~\cite{Tieckethessis}  \\
$^{6}$Li-$^{85}$Rb   & 8.87   &    -14.88   &   2450${^\flat}$  &   ~\cite{Madison2010Li6Rb85}  \\
$^{6}$Li-$^{87}$Rb   & 0.5   &    -18.6   &   2452${^\flat}$   &   ~\cite{Dalgarno2001C6,Courteille2009Li7Rb87}  \\
$^{6}$Li-$^{133}$Cs   & 30.252   &    -34.259   &   2955${^\flat}$  &   ~\cite{Weidemuller2013Li6Cs133,Dalgarno2001C6}  \\
$^{7}$Li-$^{7}$Li  & 33.75${^\flat}$   &    -26.92   &   1393.39  &   ~\cite{Hutson2014Li7,Yan1996LiC6}  \\

\hline
\end{tabular}
\\
\label{table6}
\end{table}

\setcounter{table}{0}
\renewcommand\thetable{VIII}
\begin{table}[h]
\renewcommand\arraystretch{1.5}
\caption{(Continued.)}
\begin{tabular}{p{1.6cm}p{1.6cm}<{\centering}p{1.6cm}<{\centering}p{1.5cm}<{\centering}p{1.6cm}<{\centering}}
\hline
\hline
 System   &   $a_s$ (a.u.)    &   $a_t$ (a.u.)   &   $C_{6}$ (a.u.)     &  Reference    \\
\hline
$^{7}$Li-$^{23}$Na   &  5  &   21   &   1467  &   ~\cite{Tiemann2012LiNa,Dalgarno2001C6}  \\
$^{7}$Li-$^{39}$K   & 29.1   &    81.2   &   2322  &   ~\cite{Tieckethessis}  \\
$^{7}$Li-$^{40}$K   &  13.9  &    74.5   &   2322  &   ~\cite{Tieckethessis}  \\
$^{7}$Li-$^{41}$K   & -7.92   &    69.1   &   2322  &   ~\cite{Tieckethessis}  \\
$^{7}$Li-$^{85}$Rb   & 60.5   &    -51.5   &   2450${^\flat}$  &   ~\cite{Dalgarno2001C6,Courteille2009Li7Rb87}  \\
$^{7}$Li-$^{87}$Rb   & 53.9   &    -63.5   &   2448${^\flat}$  &   ~\cite{Dalgarno2001C6,Courteille2009Li7Rb87}  \\
$^{7}$Li-$^{133}$Cs   & 45.477   &    908.6   &   2955${^\flat}$  &   ~\cite{Weidemuller2013Li6Cs133,Dalgarno2001C6}  \\
$^{23}$Na-$^{23}$Na &  18.81   &     62.58${^\flat}$   &    1560.1  &   ~\cite{Oberthaler2011Na}  \\
$^{23}$Na-$^{39}$K   &  255  &   -84   &   2350${^\flat}$  &   ~\cite{Simoni2016NaK,Dalgarno2001C6}  \\
$^{23}$Na-$^{40}$K   &  63  &   -838   &   2370${^\flat}$  &   ~\cite{Simoni2016NaK,Dalgarno2001C6}  \\
$^{23}$Na-$^{41}$K   &  -3.65  &   267   &   2360${^\flat}$  &   ~\cite{Simoni2016NaK,Dalgarno2001C6}  \\
$^{23}$Na-$^{85}$Rb   &  396  &   81   &   2472${^\flat}$   &   ~\cite{Tiemann2005NaRb}  \\
$^{23}$Na-$^{87}$Rb   &  109  &   70   &   2472${^\flat}$   &   ~\cite{Tiemann2005NaRb}  \\
$^{23}$Na-$^{133}$Cs   &  513  &   33   &   3035${^\flat}$  &   ~\cite{Tiemann2006NaCs}  \\
$^{39}$K+$^{39}$K & 138.9   &    -30.1${^\flat}$   &   3710${^\flat}$  &   ~\cite{Simoni2007K39}  \\
$^{39}$K+$^{40}$K   &  -2.84  &   -1985   &   3925.9  &   ~\cite{Tiecke2011PropertiesOP}  \\
$^{39}$K+$^{41}$K   &  113.07  &   177.1   &   3925.9  &   ~\cite{Tiecke2011PropertiesOP}  \\
$^{39}$K+$^{85}$Rb   &  33.4  &   63.9   &   4150${^\flat}$  &   ~\cite{Tiemann2007KRb}  \\
$^{39}$K+$^{87}$Rb   &  1868  &   35.9   &   4085${^\flat}$  &   ~\cite{Tiemann2007KRb}  \\
$^{39}$K+$^{133}$Cs   &  -18.4  &   70${^\flat}$   &   5159  &   ~\cite{Hutson2017KCs,Dalgarno2001C6}  \\
$^{40}$K+$^{40}$K  & 101.8${^\flat}$   &    169.67   &   3925.9  &   ~\cite{Tiecke2011PropertiesOP}  \\
$^{40}$K+$^{41}$K   &  -54.28  &   94.95${^\flat}$   &   3925.9  &   ~\cite{Tiecke2011PropertiesOP}  \\
$^{40}$K+$^{85}$Rb   &  65.8  &   -28.55   &   4150${^\flat}$  &   ~\cite{Tiemann2007KRb}  \\
$^{40}$K+$^{87}$Rb   &  -111.5  &   -215.6   &   4150${^\flat}$  &   ~\cite{Tiemann2007KRb}  \\
$^{40}$K+$^{133}$Cs   &  -51.44  &   -71.67   &   5159  &   ~\cite{Hutson2017KCs,Dalgarno2001C6}  \\
$^{41}$K+$^{41}$K & 85.53   &    58.89   &   3925.9  &   ~\cite{Yao2017dwave,Tiecke2011PropertiesOP}  \\
$^{41}$K+$^{85}$Rb   &  103.1  &   349.8   &   4150${^\flat}$  &   ~\cite{Tiemann2007KRb}  \\
$^{41}$K+$^{87}$Rb   &  7.06  &   164.4   &   4150${^\flat}$  &   ~\cite{Tiemann2007KRb}  \\
$^{41}$K+$^{133}$Cs   &  -72.79  &   179.06   &   5159  &   ~\cite{Hutson2017KCs,Dalgarno2001C6}  \\
$^{85}$Rb+$^{85}$Rb & 2735   &    -386   &   4505${^\flat}$   &   ~\cite{Cornish2013Rb85,Tiemann2010Rbpotential}  \\
$^{85}$Rb+$^{87}$Rb   &  11.37  &   184${^\flat}$   &   4710  &   ~\cite{Tiemann2010Rbpotential}  \\
$^{85}$Rb+$^{133}$Cs   &  585.6  &   11.27   &   5390${^\flat}$  &   ~\cite{Cornish2013RbCs,Tiemann2012RbCs}  \\
$^{87}$Rb+$^{87}$Rb & 90.35   &    99.04   &   4410${^\flat}$  &   ~\cite{Tiemann2010Rbpotential}  \\
$^{87}$Rb+$^{133}$Cs   &  997  &   513.3   &   5300${^\flat}$  &   ~\cite{Cornish2014Rb87Cs,Tiemann2012RbCs}  \\
$^{133}$Cs+$^{133}$Cs  & 286.5   &    2858   &   6400${^\flat}$  &   ~\cite{Grimm2013Cs}  \\
\hline
\hline
\end{tabular}
\\
\begin{flushleft}
$\flat$ The parameters are adjusted from the values given in references.
\end{flushleft}
\label{table7}
\end{table}

\FloatBarrier

%

\end{document}